\documentclass[
	aps,
	groupedaddress,
	letterpaper,
	pre,
	reprint,
	showpacs,
	twocolumn
]{revtex4-1}

\usepackage[utf8]{inputenc} % escrever os acentos diretamente ao invés de usar macros - http://aleteia.wordpress.com/2008/07/30/latex_e_acentos/
\usepackage[usenames,dvipsnames]{xcolor} % make use of these color features the color package must be inserted into the preamble. (\textcolor{red}{text})
\usepackage{natbib} % fazer citações - A short guide to BibTex: http://www.economics.utoronto.ca/osborne/latex/BIBTEX.HTM
\usepackage{amsmath} % provides various features to facilitate writing math formulas and to improve the typographical quality of their output
\usepackage{units}  % para, entre outras coisas, usar o \nicefrac
\usepackage{graphicx} % inserir graficos

\usepackage{pstricks} %
\usepackage{siunitx} %

\begin{document}

%%%%% TITLE PAGE
\title{Minimal Fragmentation of Regular Polygonal Plates}

\author{Laércio Dias}
\author{Fernando Parisio}
\affiliation{Departamento de Física, Universidade Federal de Pernambuco, 50670-901 Recife, Pernambuco, Brazil}

\begin{abstract}
Minimal fragmentation models intend to unveil the statistical properties of large ensembles of identical objects, each one 
segmented in {\it two} parts only. Contrary to what happens in the multifragmentation of a single body, minimally fragmented 
ensembles are often amenable to analytical treatments, while keeping key features of multifragmentation.
In this work we present a study on the minimal fragmentation of regular polygonal plates with up to $100$ sides. 
We observe in our model the typical statistical behavior of a solid teared apart by a strong impact, for example.
That is to say, a robust power law, valid for several decades, in the small mass limit. In the present case we were able
to analytically determine the exponent of the accumulated mass distribution to be $\nicefrac{1}{2}$. Less usual, but also
reported in a number of experimental and numerical references on impact fragmentation, is the presence of a sharp crossover 
to a second power-law regime, whose exponent we found to be $\nicefrac{1}{3}$ for an isotropic model and $\nicefrac{2}{3}$ 
for a more realistic anisotropic model.

\end{abstract}


\maketitle

%%%%%%%%%%%%%%%%%%%%%%%%%%%%%%%%%%%%%%%%%%%%%%%%%%%%%%%%%%%%%%%%%%%%%%%%%%%%%%%%%%%%%%%%%%%%%%%%%%%
% INTRODUCTION
%%%%%%%%%%%%%%%%%%%%%%%%%%%%%%%%%%%%%%%%%%%%%%%%%%%%%%%%%%%%%%%%%%%%%%%%%%%%%%%%%%%%%%%%%%%%%%%%%%%

\section{Introduction}
\label{sec:introduction}

Multifragmentation of solids in its various forms \cite{grady,Belmont, villermaux} is amongst the toughest problems in the physics of complex systems, especially regarding the prospects to reach closed analytical results of some generality. Very few statistical fragmentation models are amenable to a fully analytical approach, among them, the random fragmentation of a line \cite{lienau}, the model by Mott to describe the fragmentation of a flat surface into rectangles with random side lengths \cite{mott} (for a recent account see \cite{mott2}), and the minimal fragmentation of an ensemble of rectangular plates \cite{ParisioDias2011}. Although all these constructions in the realm of geometrical probability \cite{kendal} are highly idealized, they do shed some light into more realistic aspects of multifragmentation problems, for instance, the existence of power-law regimes in the fragment size distribution in the limit of small mass~\cite{ParisioDias2011}. 

The minimal fragmentation (MF) model of planar objects consists in considering a large collection of identical bodies split in two fragments only, instead of a single body cracked in a large number of pieces~\cite{ParisioDias2011}.
In our minimal fragmentation model for a polygon a crack is represented by a straight segment that is fully characterized by two random variables: $l\in [0,L]$ representing one of the crack limits intersecting the border of the plate, with $L$ being the polygon perimeter, and, $\phi\in[-\pi/2,\pi/2]$ being the angle between the segment and the normal direction to the side selected by the variable $l$. Initially we don't consider any directional or positional bias such that $l$ and $\phi$, have uniform distributions. Other situations can be considered, as for example, breaking isotropy by selecting a preferred direction for the crack, which amounts to a non-uniform distribution for $\phi$. We will return to this point later.

In this work we consider the MF of a regular polygon with total mass $M$ uniformly distributed over its surface. Our objective is to study the most commonly recorded quantity in experiments and simulations, namely, the distribution of fragment masses. For the sake of mathematical convenience our reasoning will be in terms of the complementary accumulated probability $\mathcal{P}(m) = 1 - \mathcal{P}_{>}(m)$, that is
\begin{equation*}
\mathcal{P}(m) = \int_{0}^{m} dm^{\prime}\;p(m^{\prime}),
\end{equation*}
where $p(m^{\prime})$ is the probability density function and $\mathcal{P}_{>}(m)$ is the probability to find a fragment with mass larger than $m$ (more usual in experimental papers). 
The article is organized as follows: in the next section we give a few preliminary definitions and in Section~\ref{sec:triangle} we start with the MF of triangular plates. There we derive in detail exact results for $\mathcal{P}(\mu)$. In section~\ref{sec:square} we review some results found in~\cite{ParisioDias2011} on the MF of squared plates. Section~\ref{sec:circle} presents a study of circular plates minimally fragmented. In Section~\ref{sec:polygon} the case of regular polygons with an arbitrary number of side is considered. In Section~\ref{sec:cosine} we introduce anisotropy in the model. Our main conclusions are summarized in Section~\ref{sec:conclusion}.

\section{Preliminary definitions}
We will denote the complementary accumulated probability associated to an ensemble of $n$-sided regular polygons by $\mathcal{P}^{(n)}(m)$.
For a fixed point in its perimeter, i.\ e.,  for a fixed value of $l$, it is useful to define the auxiliary conditional distribution $\mathcal{P}^{(n)}(m|l)$, such that,
\begin{equation}
\label{eq:Pusandolfixo}
\mathcal{P}^{(n)}(m) = \frac{1}{L}\int_{0}^{L} dl\;\mathcal{P}^{(n)}(m|l),\quad 0\leq m \leq M.
\end{equation}
It is clear, therefore, that $\mathcal{P}^{(n)}(m|l)$ stands for the conditional probability to get a fragment with mass smaller than $m$ out of the sub-ensemble in which all the cracks started at the same point on the polygon's perimeter.
 
An important feature in MF is that only two fragments are generated per event. For each fragment of mass $m$, there is another one with mass $M-m$, and, thus, $\mathcal{P}^{(n)}(m) = \mathcal{P}^{(n)}_{>}(M-m)=1-\mathcal{P}^{(n)}(M-m)$. In particular, for $m=M/2$ we have $\mathcal{P}^{(n)}(m) = \mathcal{P}^{(n)}(M-m) = \nicefrac{1}{2}$. This implies that all information can be captured by taking into account only the smallest fragment for each event in the ensemble. Thus, without loss of generality, one can work with the normalized mass $\mu=2m/M$, where $0\leq m \leq M/2$ or, equivalently, $0\leq \mu \leq 1$, with $\mathcal{P}^{(n)}(\mu=1)=1$. We get
\begin{equation}
\label{eq:PusandolfixoNorm}
\mathcal{P}^{(n)}(\mu) = \frac{1}{L}\int_{0}^{L} dl\;\mathcal{P}^{(n)}(\mu|l),\quad 0\leq \mu \leq 1,
\end{equation}
which, of course, presents the same properties of (\ref{eq:Pusandolfixo}).
%
%
%
%
%
%%%%%%%%%%%%%%%%%%%%%%%%%%%%%%%%%%%%%%%%%%%%%%%%%%%%%%%%%%%%%%%%%%%%%%%%%%%%%%%%%%%%%%%%%%%%%%%%%%%
% TRIANGLE
%%%%%%%%%%%%%%%%%%%%%%%%%%%%%%%%%%%%%%%%%%%%%%%%%%%%%%%%%%%%%%%%%%%%%%%%%%%%%%%%%%%%%%%%%%%%%%%%%%%

\section{Triangle}
\label{sec:triangle}

\begin{figure}[!h]
\centering
\includegraphics[width=0.3\textwidth]{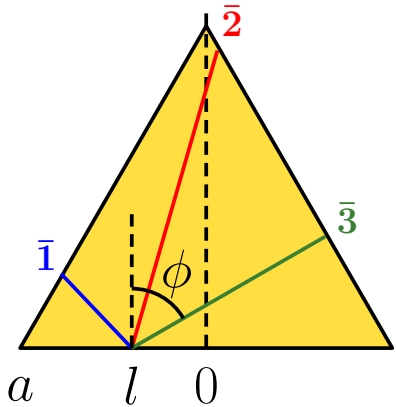}
\caption{(Color online) Three minimal fragmentation events represented on the same equilateral triangle. Fragments can be either triangular or trapezoidal. The small fragments are in the left of crack ($\bar{1}$) and ($\bar{2}$) and in the right of crack ($\bar{3}$), respectivelly.}
\label{fig:triangle}
\end{figure}

Let us begin with the simplest polygon in Euclidian geometry. Consider an equilateral triangle with perimeter $L=6a$ and uniform mass distribution. To deal with its MF, firstly, we see that it is sufficient to restrict the variable $l$ to the interval $[0,2a]$, that is, to integrate (\ref{eq:PusandolfixoNorm}) over one of the equivalent sides. In addition, note that all possible shapes resulting from the MF of an equilateral triangle are schematically described in Figure \ref{fig:triangle} by fragments of type $\bar{1}$ (triangles), type $\bar{2}$ (trapezoids), or type $\bar{3}$ (triangles), which makes it clear that, because of the reflection symmetry over the triangle heights, in fact, we only need to consider the interval $[0,a]$, with $6$ multiplying the final integral. Note carefully the difference between type $\bar{1}$ and type $\bar{3}$ fragments. In the former the smaller fragment is located at the left-hand side, while the latter is in the right-hand side.  Equation (\ref{eq:PusandolfixoNorm}) becomes
\begin{equation}
\label{eq:PusandolfixoNorm2}
\mathcal{P}^{(3)}(\mu) = \frac{1}{a}\int_{0}^{a} dl\;\mathcal{P}^{(3)}(\mu|l),\quad 0\leq \mu \leq 1,
\end{equation}

Since we are taking $\phi$ as a uniform random variable, $\mathcal{P}^{(3)}(\mu|l)=\Delta\phi/\pi$, where $\Delta\phi$ stands for the angular interval for which the fragment mass is smaller than $\mu$. Let us calculate $\mathcal{P}^{(3)}(\mu|l)$ in detail for type $\bar{1}$ fragments. First note that the maximum mass (non-normalized) is $\sqrt{3}a(a-l)/2$, implying that, for a fixed value of $l$, $\mu\leq 1-l/a$. Equivalently, if we fix a value for $\mu$, we get $l\leq a(1-\mu)$. If a realization of the random variable $\phi$ is such that a fragment type $\bar{1}$ is produced, then $\mu=(1-l/a)^2/(1-\sqrt{3}\tan\phi_{lim})$, where $\phi_{lim}$ is the angle for which the fragment mass is exactly $\mu$. We have $\phi_{lim}=\tan^{-1}[(1/\sqrt{3} - (1-l/a)^2/\sqrt{3}\mu)]$. Therefore, for fragments type $\bar{1}$ we obtain $\Delta\phi=\phi_{lim}-(-\pi/2)$, which leads to
\begin{equation}
\label{eq:Plm1}
\mathcal{P}^{(3)}_{\bar{1}}(\mu|l) = \frac{1}{\pi}\left\{\frac{\pi}{2}+\tan^{-1}\left[\frac{1}{\sqrt{3}} - \frac{(1-l/a)^2}{\sqrt{3}\mu}\right]\right\}.
\end{equation}
Similarly, for fragments of type $\bar{2}$ we have $\mu\in[1-l/a,1]$ for a fixed value of $l$, or, $l\in[a(1-\mu),a]$ for a fixed value of $\mu$. The normalized mass is given by $\mu=2-(1+l/a)^2/(1+\sqrt{3}\tan\phi_{lim})$, yielding
\begin{equation}
\label{eq:Plm2}
\mathcal{P}^{(3)}_{\bar{2}}(\mu|l) = \frac{1}{\pi}\left\{\frac{\pi}{2}+\tan^{-1}\left[\frac{(1+l/a)^2}{\sqrt{3}(2-\mu)} - \frac{1}{\sqrt{3}}\right]\right\},
\end{equation}
where $\Delta\phi=\phi_{lim}-(-\pi/2)$.
For fragments of type $\bar{3}$ we have $0\leq\mu\leq 1$ and $0\leq l \leq a$. They will occur for non-negative values of $\phi$ in the interval $[\phi_{lim}, \pi/2]$, where $\phi_{lim}$ is the angle corresponding to $\mu=1$, hence $\Delta\phi=\pi/2-\phi_{lim}$. For allowed values of $l$ and $\phi_{lim}$ we get a normalized mass given by $\mu=(1+l/a)^2/(1+\sqrt{3}\tan\phi_{lim})$, and consequently
\begin{equation}
\label{eq:Plm3}
\mathcal{P}^{(3)}_{\bar{3}}(\mu|l) = \frac{1}{\pi}\left\{\frac{\pi}{2} - \tan^{-1}\left[\frac{(1+l/a)^2}{\sqrt{3}\mu} - \frac{1}{\sqrt{3}}\right]\right\}.
\end{equation}
Gathering together expressions (\ref{eq:Plm1}), (\ref{eq:Plm2}), and (\ref{eq:Plm3}), the accumulated probability (\ref{eq:PusandolfixoNorm2}) becomes
\begin{equation*}
\begin{split}
\label{eq:PusandolfixoNorm3}
\mathcal{P}^{(3)}(\mu) = \frac{1}{a} \left[ \int_{0}^{a(1-\mu)} dl\;\mathcal{P}^{(3)}_{\bar{1}}(\mu|l) + \int_{a(1-\mu)}^{a} dl\;\mathcal{P}^{(3)}_{\bar{2}}(\mu|l) \right. \\
{}+ \left. \int_{0}^{a} dl\;\mathcal{P}^{(3)}_{\bar{3}}(\mu|l) \right],
\end{split}
\end{equation*}
which can be integrated and, after some algebra, results in 
\begin{align}
\label{eq:P3m}
\mathcal{P}^{(3)}(\mu) =\;&{}1 - \frac{1}{2\pi}\left\{
12\tan^{-1}\left(\frac{1-\mu}{\sqrt{3}}\right) \right. \nonumber\\
&- \sqrt{6\mu}\left[\pi - 2\tan^{-1}\left(\frac{\sqrt{2\mu}}{2-\mu}\right)\right] \\
&+ \left. \sqrt{6(2-\mu)}\left[\pi - 2\tan^{-1}\left(\frac{\sqrt{2(2-\mu)}}{\mu}\right)\right] 
\right\}.\nonumber
\end{align}
All information about the ensemble is contained in the above relation.

We will refer to the limit of very small fragment mass as the dust regime. The result found in (\ref{eq:P3m}) implies that $\mathcal{P}^{(3)}(\mu)$ behaves as a power-law in this limit. Explicitly,
\begin{equation}
\label{eq:dustTriangle}
\mathcal{P}^{(3)}(\mu) \approx \sqrt{\frac{3}{2}}\;\mu^{\nicefrac{1}{2}},\quad\mbox{for }\mu\rightarrow 0.
\end{equation}
We leave a more detailed discussion on this behavior to a later section, after we have presented our results for general regular polygons.
%
%%%%%%%%%%%%%%%%%%%%%%%%%%%%%%%%%%%%%%%%%%%%%%%%%%%%%%%%%%%%%%%%%%%%%%%%%%%%%%%%%%%%%%%%%%%%%%%%%%%
% SQUARE
%%%%%%%%%%%%%%%%%%%%%%%%%%%%%%%%%%%%%%%%%%%%%%%%%%%%%%%%%%%%%%%%%%%%%%%%%%%%%%%%%%%%%%%%%%%%%%%%%%%

\section{Square}
\label{sec:square}

In this section we will recast some analytical results previously obtained in~\cite{ParisioDias2011}. We consider this review to be necessary in order to clarify our procedure in the case of regular polygons with arbitrary number of sides. In Equation (9) of~\cite{ParisioDias2011} the mass distribution for rectangles of arbitrary aspect ratios $\gamma$ is given within the same MF model. To obtain the corresponding distribution $\mathcal{P}^{(4)}(\mu)$ for an ensemble of squares we simply set $\gamma=1$. The obtained expression, after some trigonometric simplification, can be written as
\begin{align}
\label{eq:P4m}
\mathcal{P}^{(4)}(\mu) = &\frac{2}{\pi}(\mu+1)\tan^{-1}(\mu) \nonumber\\
&+ \frac{\sqrt{2\mu}}{\pi}\left\{\pi - 2\tan^{-1}\left( \frac{\sqrt{2\mu}}{1-\mu} \right) \right\}.
\end{align}

\begin{figure}[!h]
\centering
\includegraphics[width=0.3\textwidth]{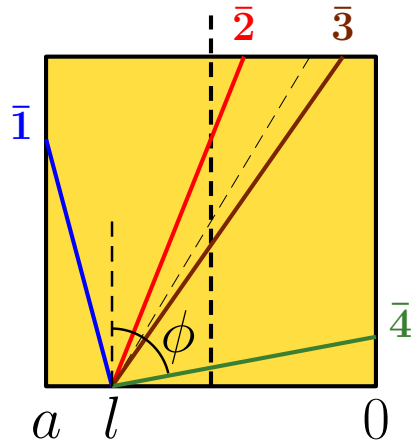}
\caption{(Color online) MF of a square. Small fragment can have geometrical forms according to cracks of type ($\bar{1}$),  ($\bar{2}$),  ($\bar{3}$), or ($\bar{4}$). Smaller fragments are in the left-hand side in the two first cases and in the right-hand side in the two last cases.}
\label{fig:square}
\end{figure}

For the sake of clarity, let us derive the above expression by employing the same ideas of the previous section. In Figure~\ref{fig:square} we present all possible fragment geometries: type $\bar{1}$ and $\bar{4}$ are triangles, while type $\bar{2}$ and $\bar{3}$ are trapezoids. Reflexion upon the vertical symmetry axis makes type $\bar{1}$ fragments become type $\bar{4}$ fragments, as well, type $\bar{2}$ fragments are turned into type $\bar{3}$ fragments. Thus, if for each value of $l$, we consider only fragments of type $\bar{1}$ and $\bar{2}$, we have exactly half of the total number. Therefore,
\begin{equation}
\label{eq:PusandolfixoNorm4}
\mathcal{P}^{(4)}(\mu) = \frac{2}{a} \left[ \int dl\;\mathcal{P}^{(4)}_{\bar{1}}(\mu|l) + \int dl\;\mathcal{P}^{(4)}_{\bar{2}}(\mu|l) \right]
\end{equation}

For type $\bar{1}$ fragments and a fixed value of $l$, $\mu \leq 1 - l/a$, or equivalently, for a fixed value of $\mu$, $l \leq a(1-\mu)$. For these fragments the normalized mass is given by $\mu = - (1-l/a)^{2}/\tan(\phi)$. From the last relation we obtain $\phi = \phi(\mu)$ and $\Delta\phi_{\bar{1}}$. By the same token, for fragments of type $\bar{2}$, $\mu \in [a(1-\mu), a]$ and $\mu = 2(1-l/a)+\tan(\phi)$. Thus, we have
\begin{align}
\label{eq:PusandolfixoNorm4-2}
\mathcal{P}^{(4)}(\mu) =\;&{}\frac{2}{a}\left\{
\int_{0}^{a(1-\mu)} \frac{1}{\pi} \left[ \frac{\pi}{2} - \tan^{-1}\left(\frac{(1-l/a)^2}{\mu}\right) \right] dl \right. \nonumber\\
&+ \left. \int_{a(1-\mu)}^{a} \frac{1}{\pi} \left[ \frac{\pi}{2} + \tan^{-1}(\mu - 2(1-l/a)) \right] dl \right\},\nonumber
\end{align}
which after integration yields~(\ref{eq:P4m}).

The dust-regime power law is found from~(\ref{eq:P4m}) by expanding around $\mu=0$, which results in
\begin{equation}
\label{eq:dustSquare}
\mathcal{P}^{(4)}(\mu) \approx \sqrt{2}\;\mu^{\nicefrac{1}{2}},\quad\mbox{for }\mu\rightarrow 0,
\end{equation}
which, apart from the multiplicative constant, coincides with the result for the triangle.

The next case one should address would be that of a regular pentagon. However, the modest $n=5$ is already almost prohibitive 
in terms of analytical calculations, and the difficulty quickly increases with $n$. Thus, for $n \ge 5$, our approach will be mainly numerical. 
There are, however, some partial analytical results that can be obtained in the general case. For simplicity, in the next section we jump to the most symmetrical case, corresponding to the limit $n \rightarrow \infty$.

%
%
%
%
%
%%%%%%%%%%%%%%%%%%%%%%%%%%%%%%%%%%%%%%%%%%%%%%%%%%%%%%%%%%%%%%%%%%%%%%%%%%%%%%%%%%%%%%%%%%%%%%%%%%%
% CIRCLE
%%%%%%%%%%%%%%%%%%%%%%%%%%%%%%%%%%%%%%%%%%%%%%%%%%%%%%%%%%%%%%%%%%%%%%%%%%%%%%%%%%%%%%%%%%%%%%%%%%%
\section{Circle}
\label{sec:circle}

\begin{figure}[!h]
\centering
\includegraphics[width=0.3\textwidth]{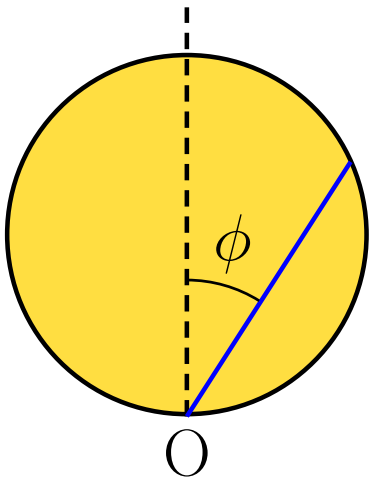}
\caption{(Color online) Minimal fragmentation of a disk. Due to the continuous symmetry of the circle one only has to consider a single perimeter point.}
\label{fig:circle}
\end{figure}

In the limit $n\rightarrow \infty$ the polygon becomes a circle, the crack being a chord linking two perimeter points. Due to the continuous rotational symmetry, we can always admit that one of these points is fixed. Let $O$ be this fixed point. Thus, $\phi$ represents the angle between the diameter line containing $O$ and the crack (see Figure~\ref{fig:circle}).

\begin{figure}[!h]
  \centering
    \includegraphics[width=0.45\textwidth]{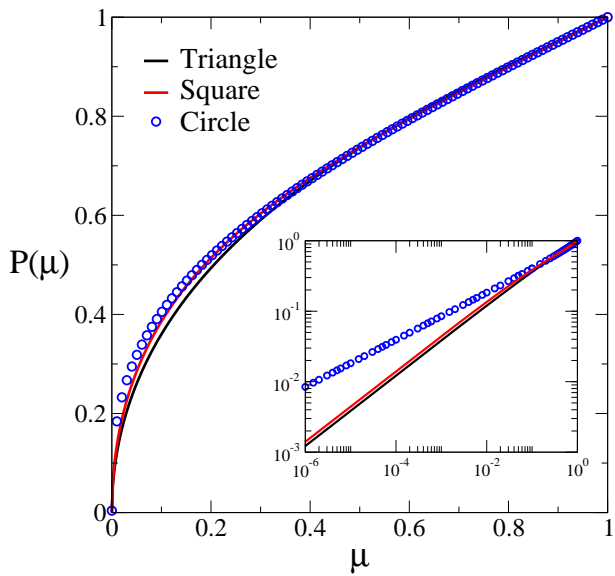}
\caption{``Shapes'' of the accumulated probabilities for triangle, square and circle minimal fragmentation. The inset depicts the same functions
in a loglog plot.}
\label{fig:P-3-4-infty}
\end{figure}

Reflection symmetry about the diameter implies that we just need to consider $0 \leq \phi \leq \pi/2$. Observe that $\phi=0$ produces $\mu=1$ whereas $\phi=\pi/2$ produces $\mu=0$. Since $\phi$ is a random variable with uniform distribution and, in principle, we can obtain $\phi=\phi(\mu)$, we get the complementary accumulated probability, that reads
\begin{equation}
\label{eq:Pcircle1}
\mathcal{P}(\mu) = \frac{\Delta\phi}{\pi/2} = \frac{\pi/2 - \phi(\mu)}{\pi/2},
\end{equation}
where we need to obtain $\phi(\mu)$. However $\mu(\phi)$ is given by
\begin{equation}
\label{eq:muCircle}
\mu = 1 - \frac{2}{\pi}\bigg[\phi + \cos\phi\sin\phi\bigg],
\end{equation}
which turns out not to be invertible due to its transcendental nature. One can solve the equation numerically and find the curve that represents the mass distribution. Also, we can simulate a few thousands of MF events and estimate the accumulated probability. The result of such approach is shown in Figure~\ref{fig:P-3-4-infty} together with plots of (\ref{eq:P3m}) and (\ref{eq:P4m}). Although the shapes of these curves are almost indistinguishable in a cartesian plot, an equivalent loglog plot shows that there are quite distinct power laws involved. 

In this regard, we can get more precise information, at least in the dust regime, from (\ref{eq:muCircle}). Here, small masses are equivalent to $\phi\approx\pi/2\;\si{\radian}$, that is
\begin{equation}
\mu \approx \frac{4}{3\pi}\bigg(\frac{\pi}{2}-\phi\bigg)^{3},
\label{eq:circleSmallMass}
\end{equation}
to lowest non-vanishing order.
Replacing this approximation in (\ref{eq:Pcircle1}), again, we obtain a power-law
\begin{equation}
\label{eq:Pncircle}
\mathcal{P}(\mu) \approx \bigg(\frac{6}{\pi^2}\bigg)^{\nicefrac{1}{3}}\mu^{\nicefrac{1}{3}},\quad\mbox{for }\mu\rightarrow 0.
\end{equation}
Interestingly enough, this time, the exponent is $\nicefrac{1}{3}$.
This result leaves us with two extrapolation hypotheses for arbitrary $n$ that stand out as, arguably, the most reasonable ones. 
Either, (i) although the triangle and the square present the same exponent in the dust regime, starting from the pentagon,
the exponent continuously decreases until its asymptotic value of $\nicefrac{1}{3}$ for $n \rightarrow \infty$, or (ii) the exponent for all regular polygons with a finite number of sides is $\nicefrac{1}{2}$, becoming $\nicefrac{1}{3}$ only for the circle. Hypothesis (i) might sound strange since the dust regime exponent does not change from $n=3$ to $n=4$. Yet, this is not so unusual. Consider, e.\ g., the Gamma function $\Gamma(n)$ that assumes the same value for $n=1$ and $n=2$, and then becomes an increasing function of $n$ for $n>2$. As for the second hypothesis, it seems to suggest a discontinuity in the limit $n \rightarrow \infty$. As we will see in the next section, this issue is a sensitive one and must be handled with care. Indeed, we will show that supposition (i) is wrong and, although (ii) is correct, it does not tell the whole story.

%
%
%
%
%
%%%%%%%%%%%%%%%%%%%%%%%%%%%%%%%%%%%%%%%%%%%%%%%%%%%%%%%%%%%%%%%%%%%%%%%%%%%%%%%%%%%%%%%%%%%%%%%%%%%
% POLYGON
%%%%%%%%%%%%%%%%%%%%%%%%%%%%%%%%%%%%%%%%%%%%%%%%%%%%%%%%%%%%%%%%%%%%%%%%%%%%%%%%%%%%%%%%%%%%%%%%%%%
\section{Arbitrary Regular Polygon}
\label{sec:polygon}

As we did for the MF of a circle, we can also find the exponent associated to the dust regime analytically for an arbitrary polygon. Indeed, notice that this regime must come exclusivelly from triangular fragments (type $\bar{1}$ or type $\bar{n}$ generalizing Figure~\ref{fig:square}), because only these fragments can have vanishingly small masses, leading to an expression analogous to (\ref{eq:PusandolfixoNorm4}), however, without the second term. Thus, we need to evaluate
\begin{equation}
\label{eq:PusandolfixoNorm_n}
\mathcal{P}^{(n)}(\mu) \approx \frac{2}{a} \int dl\;\mathcal{P}^{(n)}_{\bar{1}}(\mu|l).
\end{equation}

\noindent For a fixed value of $l$, all type $\bar{1}$ fragments comply with
\begin{equation*}
\label{eq:muFixedl}
\mu \leq \frac{4}{n}\left(1 - \frac{l}{a}\right)\tan\left(\frac{\pi}{n}\right)\sin\left(\frac{2\pi}{n}\right),
\end{equation*}
or, for a fixed $\mu$, $l \leq a(1-f_{n}\mu)$, where
\begin{equation*}
\label{eq:fn}
f_{n} = \frac{n}{4\tan(\pi/n)\sin(2\pi/n)},
\end{equation*}
for any finite $n$. So, integration limits are defined. Now, we recall that $\Delta\phi/\pi = (\phi_{lim} - (-\pi/2))/\pi$. In the present case
\begin{equation*}
\label{eq:phiLim}
\phi_{lim} = -\tan^{-1} \left[a_n + \frac{(1-l/a)^2}{b_n\mu}\right],
\end{equation*}
where $a_n = \cot(2\pi/n)$ and $b_n^{-1} = (4/n)\tan(\pi/n)$. Gathering all these elements in~(\ref{eq:PusandolfixoNorm_n}), we get
\begin{align}
\label{eq:Pn1}
\mathcal{P}^{(n)}(\mu) &\approx 1 - f_n\mu \nonumber\\
&- \frac{2}{a\pi} \int_{0}^{a(1-f_{n}\mu)} dl\;\tan^{-1} \left[a_n + \frac{(1-l/a)^2}{b_n\mu}\right].
\end{align}
For $\mu\approx 0$ one can write
\begin{align}
\label{eq:integration}
\int_{0}^{a(1-f_{n}\mu)} dl\;\tan^{-1} [\ldots] =&\; \int_{0}^{a} dl\;\tan^{-1} [\ldots] \nonumber\\
&- a f_n \mu \tan^{-1} [\ldots].\nonumber
\end{align}
Hence, we get $\mathcal{P}^{(n)}(\mu) \approx \sqrt{\frac{n}{2}}\;\mu^{\nicefrac{1}{2}} - c_n\;\mu + O(\mu^2)$,
where
\begin{equation*}
\label{eq:cn}
c_n = f_n \left\{ 1 + \frac{2}{\pi}\left[\sin\left(\frac{2\pi}{n}\right) - \tan^{-1}\left(\cot\left(\frac{2\pi}{n}\right)\right) \right] \right\}.
\end{equation*}
Therefore, to the lowest order we obtain
\begin{equation}
\label{eq:PnFinal}
\mathcal{P}^{(n)}(\mu) \approx \sqrt{\frac{n}{2}}\;\mu^{\nicefrac{1}{2}},\quad\mbox{for }\mu\rightarrow 0.
\end{equation}
This result shows unequivocally that hypothesis (ii) in the previous section is correct: all regular polygons with a finite number of sides present an exponent of $\nicefrac{1}{2}$ in the dust regime.  

An important point here is the distinction between the rather mathematical limit $\mu \rightarrow 0$ and the limit of small masses that can be actually accessed by numerical simulations or experimentation. 
In this regard, we remark that if we look at a regular polygon with, say $n=20$, from a modest distance, it will probably appear to be a perfectly smooth disk.
This observation suggests that, although the exponent for the dust regime ($\mu \rightarrow 0$) is $\nicefrac{1}{2}$, the exponent $\nicefrac{1}{3}$ should also appear for values of $\mu$ above some threshold (still satisfying $\mu<<1$), characterizing a crossover in the mass distribution. 

\begin{figure}[!h]
  \centering
    \includegraphics[width=0.45\textwidth]{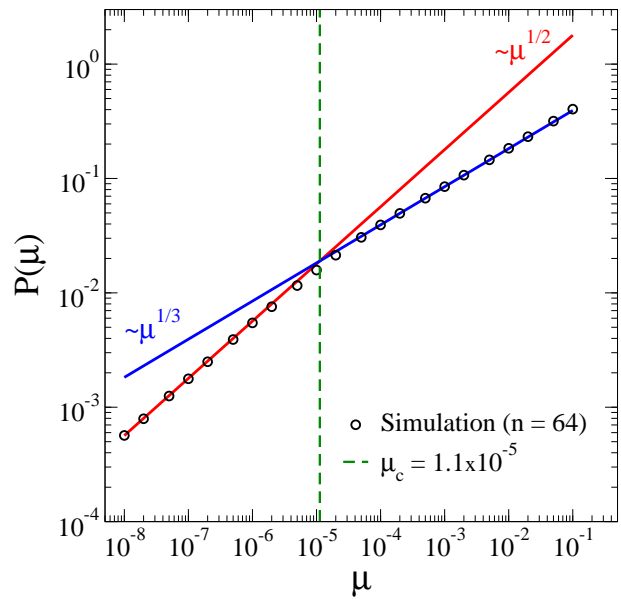}
\caption{(color online)  Simulation results for the dust-regime ($\sim \mu^{1/2}$ for $\mu<\mu_c$) and for the ``disk'' regime ($\sim \mu^{1/3}$ for $1>>\mu>\mu_c$) for $n=64$.}
\label{fig:crossover}
\end{figure}

Our numerical simulations corroborate the occurrence of this behavior.
We produced $10^7$ fragmentation events for each polygon in the range $n~\in~\{3,\ldots,100\}$. In each event the fragment area (mass) was calculated and recorded, and, the first exponent is estimated for values below $\mu_c$ (the crossover mass) and  above $\mu_{lim}$, conveniently chosen as we discuss in what follows. The second exponent is calculated for $\mu>\mu_c$. The result for the regular polygon with 64 sides is displayed in Figure~\ref{fig:crossover}, where the crossover at $\mu=\mu_c$ (vertical line) is evident, each power-law regime being valid for several decades. Actually, power laws (\ref{eq:Pncircle}) and (\ref{eq:PnFinal}) are very good approximations for the distribution $\mathcal{P}^{(n)}(\mu)$ in the range of values of $\mu<0.1$, beyond the upper values of the dust regime. In fact, one can precisely determine the crossover mass by seeking the point where the curves (\ref{eq:Pncircle}) and (\ref{eq:PnFinal}) coincide:
\begin{equation*}
\sqrt{\frac{n}{2}}\;\mu_c^{\nicefrac{1}{2}} = \bigg(\frac{6}{\pi^2}\bigg)^{\nicefrac{1}{3}}\mu_c^{\nicefrac{1}{3}},
\end{equation*}
from which we obtain
\begin{equation}
\label{eq:crossoverMass}
\mu_c(n) = \frac{288}{\pi^4}\;n^{-3}\approx 2.97\;n^{-3}.
\end{equation}

We see, therefore, that the crossover mass becomes smaller as the number of sides increases, becoming zero for $n\rightarrow \infty$ (no crossover for the disk).
For $n=15$, e.\ g., $\mu_c \approx 9 \time 10^{-4}\approx 1.5 \times 10^{-4}$, and the dust regime would hardly be observed in a hypothetical experiment. In our statistical analysis, to capture the dust regime, one has to satisfy $\mu_{lim}<\mu_c$. For the last polygon we considered ($n=100$) $\mu_{lim}<10^{-6}$.
%
%
%
%
%
%%%%%%%%%%%%%%%%%%%%%%%%%%%%%%%%%%%%%%%%%%%%%%%%%%%%%%%%%%%%%%%%%%%%%%%%%%%%%%%%%%%%%%%%%%%%%%%%%%%
% COSINE
%%%%%%%%%%%%%%%%%%%%%%%%%%%%%%%%%%%%%%%%%%%%%%%%%%%%%%%%%%%%%%%%%%%%%%%%%%%%%%%%%%%%%%%%%%%%%%%%%%%
\section{Anisotropy in the Fracture Direction}
\label{sec:cosine}

In many actual situations there is no a priori reason to assume isotropy in the angular distribution followed by the cracks. Suppose, for example, that a plate suffers a lateral impact perpendicular to one of its sides \cite{donangelo,donangelo2}. This situation is more likely to generate a fracture more or less parallel to the impact direction than a nearly tangential crack. Of course, by including this ingredient in our model we loose the ability to find complete analytical solutions for the mass distribution of triangles and squares. Still, we can find expressions for the dust regime in some cases. In this section we assume that $\phi$ obeys a cosine differential distribution given by $p(\phi)=\cos\phi,\;0\leq\phi\leq\pi/2$. Note that this will make the occurrence of small masses less common in comparison to the isotropic case. 

\begin{figure}[!h]
  \centering
    \includegraphics[width=0.45\textwidth]{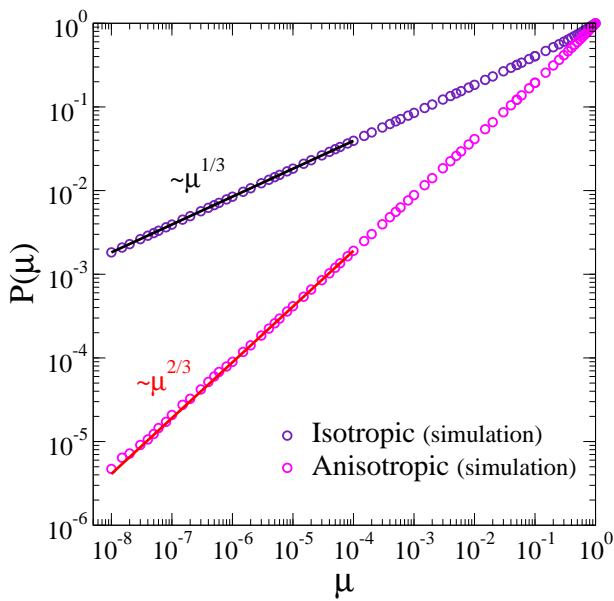}
\caption{(color online) Mass distribution for the isotropic ($\sim \mu^{1/3}$) and anisotropic ($\sim \mu^{2/3}$) MF of a disk.
Note the absense of the crossover observed for polygons.}
\label{fig:DRcircleCosine}
\end{figure}

Let us consider the MF of a circle under this new condition. The accumulated probability is $\mathcal{P}(\phi)=\sin(\phi)$. However, noting that $\phi=0$ is equivalent to $\mu=1$ and that $\phi=\pi/2$ corresponds to $\mu=0$, we see that $\mathcal{P}(\phi)$ is related to $\mathcal{P}_{>}(\mu)$ by
\begin{equation*}
\label{eq:PphiPmuEquiv1}
\mathcal{P}(\phi) = \int_{0}^{\phi}d\phi^{\prime}\;p(\phi^{\prime}) = \int_{\mu}^{1}d\mu^{\prime}\;p(\mu^{\prime}) = 1 - \mathcal{P}(\mu),
\end{equation*}
therefore,
\begin{equation*}
\label{eq:PphiPmuEquiv2}
\mathcal{P}(\mu) = 1 - \sin(\phi(\mu)).
\end{equation*}
Again, we are not able to write a closed expression for $\phi(\mu)$ because Equation~(\ref{eq:muCircle}), which also holds here, is not invertible. For very small fragments ($\phi\approx\pi/2\;\si{\radian}$), we have $\mathcal{P}(\mu(\phi)) \approx (\phi/2-\phi)^2/2$. Using Equation~(\ref{eq:circleSmallMass}) we get
\begin{equation}
\label{eq:DRcircleCos}
\mathcal{P}(\mu) \approx \bigg(\frac{3\pi}{2^{\nicefrac{7}{2}}}\bigg)^{\nicefrac{2}{3}}\mu^{\nicefrac{2}{3}},\quad\mbox{for }\mu\rightarrow 0.
\end{equation}
We, thus, obtain a power law with a larger exponent in accordance to our expectation of getting relatively less fragments with small masses. 
In Figure~\ref{fig:DRcircleCosine} we show the log-log plots of the accumulated mass distribution for the MF of a disk in the isotropic and anisotropic cases for $\mu$ between $10^{-8}$ and $1$. Note the robustness of both power laws in the first 7 decades.

\begin{figure}[!h]
  \centering
    \includegraphics[width=0.45\textwidth]{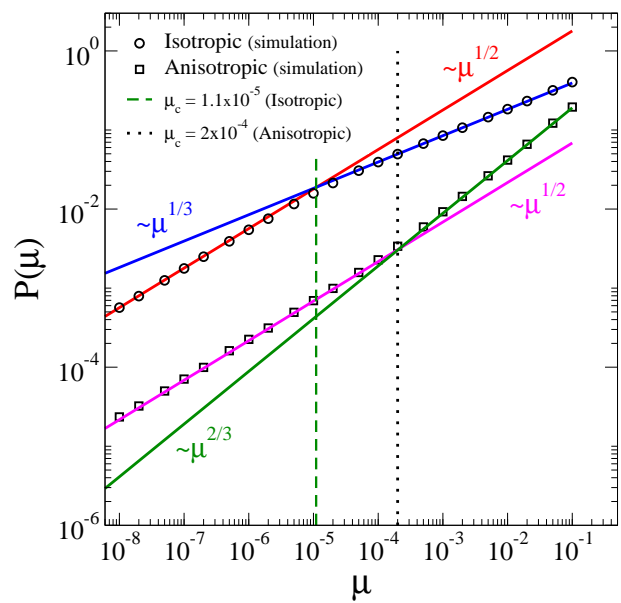}
\caption{(color online) Overall results for the MF of a polygon with $64$ sides. The crossover mass is more than ten times larger in the anisotropic scenario than its corresponding value in the isotropic case. The qualitative features displayed are fairly independent of $n$. }
\label{fig:DRpolygonCosine}
\end{figure}

Concerning the $n$-sided polygons under anisotropic MF, since $-\pi/2\leq\phi\leq\pi/2$, the probability density function is given by $p(\phi)=\cos(\phi)/2$, and $\mathcal{P}(\phi) = (1+\sin\phi)/2$. In this case we can't find closed results for $\mathcal{P}^{(n)}(\mu)$, 
even in the dust regime. However, we obtained quite convincing numerical evidence indicating that the exponent for the dust regime remains 
unchanged, the limit $\mu \rightarrow 0$ being well described by 
\begin{equation}
\label{eq:DRpolygonCos}
\mathcal{P}^{(n)}(\mu) \approx \sqrt{\frac{3}{n}}\;\mu^{\nicefrac{1}{2}},\quad\mbox{for }\mu\rightarrow 0.
\end{equation}
Given this result it is clear that the mass distribution also presents a crossover, as it happened in the isotropic case. The critical mass $\mu_c(n)$, which characterizes this crossover, is approximately given by
\begin{equation}
\label{eq:crossoverMassCosine}
\mu_c(n) \approx \frac{2^{14}}{3\pi^4}\;n^{-3}\approx 56.1 \;n^{-3}.
\end{equation}
In Figure~\ref{fig:DRpolygonCosine} we present the numerical results concerning this section in contrast with those coming from a uniform angular distribution. 
Its is perhaps reasonable to believe that the dust-regime exponent of $\nicefrac{1}{2}$ is present for a large variety of angular distributions of the fracture directions with a crossover to the exponent characterizing the fragmentation of a disk (that may change for different physical situations).
Notice that, in the anisotropic case we studied the crossover mass is more than one order of magnitude larger than its value in the isotropic model.
For $n=15$ we get $\mu_c \approx 1.7 \times 10^{-2}$, in comparison to the value obtained in the isotropic case ($1.5 \times 10^{-4}$).
Reversing the reasoning, the crossover mass related to a unknown sample may give sensitive information on the angular distribution.

%
%
%
%
%
%%%%%%%%%%%%%%%%%%%%%%%%%%%%%%%%%%%%%%%%%%%%%%%%%%%%%%%%%%%%%%%%%%%%%%%%%%%%%%%%%%%%%%%%%%%%%%%%%%%
% CONCLUSION
%%%%%%%%%%%%%%%%%%%%%%%%%%%%%%%%%%%%%%%%%%%%%%%%%%%%%%%%%%%%%%%%%%%%%%%%%%%%%%%%%%%%%%%%%%%%%%%%%%%
\section{Summary and Conclusion}
\label{sec:conclusion}

Random patterns in two dimensions are of great practical \cite{nordlund,corbett} and academic interest \cite{rivier} on the one hand, 
and, on the other hand, are easier than the analogous problems in three dimensions. Even in low-dimensional systems, multifragmentation problems are utterly complex, which, in general, hinder the possibility of obtaining information other than numeric. Minimal fragmentation models intend to provide a more tractable way to deal with, at least some features of multiple fragmentation phenomena. Also, they may be of more direct interest for other classes of problems. Consider, e.\ g., a crack propagating on a tiled floor, where typically each tile is traversed by the failure only once, thus, being minimally fragmented. In this situation the presented scheme would directly describe the observed mass distribution. In the most usual case of squared tiles we would have $\mu_c \approx 0.046$ for isotropic crack propagation allowing for a relatively easy observation of the crossover.

We have considered in detail the minimal fragmentation problem of a disk and of all regular polygons up to 100 sides. The accumulated mass distribution (number of fragments with mass smaller than a certain value) has been shown to be very well described by a composition of two 
power-law regimes. In a range of several decades of fragment masses we found that:

\begin{equation}
P(\mu) \propto
\bigg\{ 
\begin{array}{*{2}c}
\mu^{1/2} \;\;\mbox{for}\;\; \mu<\mu_c\;, \\ \mu^{\alpha}\;\; \mbox{for} \;\;\mu>\mu_c \;,
\end{array}
\end{equation}
where $\alpha=1/3$ for the isotropic model and $\alpha=2/3$ for the anisotropic model and $\mu_c \propto n^{-3}$.
The nature of this crossover is related to the fact that even a regular polygon with a few sides looks like a disk at a sufficient 
distance. Therefore, one could call it a ``proximity'' crossover. The appearance of composite power laws has been considered one of the most interesting features in the fragmentation of brittle solids and has been reported in experiments involving long thin rods \cite{matsushita} and, more explicitly, in fragmentation of plates \cite{balslev, donangelo,donangelo2} as well as in computer simulations \cite{gomes}.

The power-law divergences of themodynamical susceptibilities are a signature of criticality. In the thermodynamic limit this criticality manifests itself as a lack of characteristic scales in the onset of first order phase transitions, where fluctuations can be arbitrarily large. In fragmentation problems we are, of course, far from equilibrium and, thus, outside the realm of thermodynamics. In spite of this, the power laws we found for the MF of flat plates, although less ``critical'' in the sense that the mean value of the small fragment mass is mathematically well defined, present quite large variances.
Consider the case of the disk with inhomogeneous crack propagation. The normalized mass $\mu\approx 0.1$ is the largest mass for which (\ref{eq:Pncircle}) is valid. The average mass of small fragments is $\langle\mu \rangle \approx 0.008$, while $\Delta \mu \approx 0.031$. Thus, the root-mean square deviation is more than 30$\%$ of the whole interval $[0,0.1]$, showing that, also in this case, characteristic scales are not sharply defined.

\begin{acknowledgments}
We are indebted to José A.\ de Miranda Neto, Paulo Campos, and Sérgio Coutinho by their strong support to this project. L.\ D.\ would like to thank Laura T.\ Corredor B.\ and Pedro H.\ de Figuêiredo by their help in the preparation of figures. Financial support from the Brazilian agencies Conselho Nacional de Desenvolvimento Cient\'{\i}fico e Tecnol\'ogico (CNPq) through the program ``Instituto Nacional de Ciência e Tecnologia de Fluidos Complexos (INCT-FCx)'', Coordena\c{c}\~ao de Aperfei\c{c}oamento de Pessoal de N\'{\i}vel Superior (CAPES), and Funda\c{c}\~ao de Amparo \`a Ci\^encia e Tecnologia do Estado de Pernambuco (FACEPE) (APQ-1415-1.05/10) is acknowledged.
\end{acknowledgments}

\bibliography{referencesFragmentation}

\begin{thebibliography}{16}%
\makeatletter
\providecommand \@ifxundefined [1]{%
 \@ifx{#1\undefined}
}%
\providecommand \@ifnum [1]{%
 \ifnum #1\expandafter \@firstoftwo
 \else \expandafter \@secondoftwo
 \fi
}%
\providecommand \@ifx [1]{%
 \ifx #1\expandafter \@firstoftwo
 \else \expandafter \@secondoftwo
 \fi
}%
\providecommand \natexlab [1]{#1}%
\providecommand \enquote  [1]{``#1''}%
\providecommand \bibnamefont  [1]{#1}%
\providecommand \bibfnamefont [1]{#1}%
\providecommand \citenamefont [1]{#1}%
\providecommand \href@noop [0]{\@secondoftwo}%
\providecommand \href [0]{\begingroup \@sanitize@url \@href}%
\providecommand \@href[1]{\@@startlink{#1}\@@href}%
\providecommand \@@href[1]{\endgroup#1\@@endlink}%
\providecommand \@sanitize@url [0]{\catcode `\\12\catcode `\$12\catcode
  `\&12\catcode `\#12\catcode `\^12\catcode `\_12\catcode `\%12\relax}%
\providecommand \@@startlink[1]{}%
\providecommand \@@endlink[0]{}%
\providecommand \url  [0]{\begingroup\@sanitize@url \@url }%
\providecommand \@url [1]{\endgroup\@href {#1}{\urlprefix }}%
\providecommand \urlprefix  [0]{URL }%
\providecommand \Eprint [0]{\href }%
\providecommand \doibase [0]{http://dx.doi.org/}%
\providecommand \selectlanguage [0]{\@gobble}%
\providecommand \bibinfo  [0]{\@secondoftwo}%
\providecommand \bibfield  [0]{\@secondoftwo}%
\providecommand \translation [1]{[#1]}%
\providecommand \BibitemOpen [0]{}%
\providecommand \bibitemStop [0]{}%
\providecommand \bibitemNoStop [0]{.\EOS\space}%
\providecommand \EOS [0]{\spacefactor3000\relax}%
\providecommand \BibitemShut  [1]{\csname bibitem#1\endcsname}%
\let\auto@bib@innerbib\@empty
%</preamble>
\bibitem [{\citenamefont {Grady}\ and\ \citenamefont {Kipp}(1985)}]{grady}%
  \BibitemOpen
  \bibfield  {author} {\bibinfo {author} {\bibfnamefont {D.~E.}\ \bibnamefont
  {Grady}}\ and\ \bibinfo {author} {\bibfnamefont {M.~E.}\ \bibnamefont
  {Kipp}},\ }\href {\doibase 10.1063/1.336139} {\bibfield  {journal} {\bibinfo
  {journal} {Journal of Applied Physics}\ }\textbf {\bibinfo {volume} {58}},\
  \bibinfo {pages} {1210} (\bibinfo {year} {1985})}\BibitemShut {NoStop}%
\bibitem [{\citenamefont {Higley}\ and\ \citenamefont
  {Belmonte}(2008)}]{Belmont}%
  \BibitemOpen
  \bibfield  {author} {\bibinfo {author} {\bibfnamefont {M.}~\bibnamefont
  {Higley}}\ and\ \bibinfo {author} {\bibfnamefont {A.}~\bibnamefont
  {Belmonte}},\ }\href@noop {} {\bibfield  {journal} {\bibinfo  {journal}
  {Physica A}\ }\textbf {\bibinfo {volume} {387}},\ \bibinfo {pages} {6897}
  (\bibinfo {year} {2008})}\BibitemShut {NoStop}%
\bibitem [{\citenamefont {Vandenberghe}\ and\ \citenamefont
  {Emmanuel}(2013)}]{villermaux}%
  \BibitemOpen
  \bibfield  {author} {\bibinfo {author} {\bibfnamefont {N.}~\bibnamefont
  {Vandenberghe}}\ and\ \bibinfo {author} {\bibfnamefont {V.}~\bibnamefont
  {Emmanuel}},\ }\href@noop {} {\bibfield  {journal} {\bibinfo  {journal} {Soft
  Matter}\ }\textbf {\bibinfo {volume} {9}},\ \bibinfo {pages} {8162} (\bibinfo
  {year} {2013})}\BibitemShut {NoStop}%
\bibitem [{\citenamefont {Lienau}(1936)}]{lienau}%
  \BibitemOpen
  \bibfield  {author} {\bibinfo {author} {\bibfnamefont {C.~C.}\ \bibnamefont
  {Lienau}},\ }\href {\doibase 10.1016/S0016-0032(36)90526-4} {\bibfield
  {journal} {\bibinfo  {journal} {Journal of the Franklin Institute}\ }\textbf
  {\bibinfo {volume} {221}},\ \bibinfo {pages} {485} (\bibinfo {year}
  {1936})}\BibitemShut {NoStop}%
\bibitem [{\citenamefont {Mott}\ and\ \citenamefont {Linfoot}(1943)}]{mott}%
  \BibitemOpen
  \bibfield  {author} {\bibinfo {author} {\bibfnamefont {N.~F.}\ \bibnamefont
  {Mott}}\ and\ \bibinfo {author} {\bibfnamefont {E.~H.}\ \bibnamefont
  {Linfoot}},\ }\href@noop {} {\bibfield  {journal} {\bibinfo  {journal}
  {Ministry of Supply, Report no.\ AC3348}\ } (\bibinfo {year}
  {1943})}\BibitemShut {NoStop}%
\bibitem [{\citenamefont {Grady}(2006)}]{mott2}%
  \BibitemOpen
  \bibfield  {author} {\bibinfo {author} {\bibfnamefont {D.}~\bibnamefont
  {Grady}},\ }\href@noop {} {\emph {\bibinfo {title} {Fragmentation of Rings
  and Shells: the legacy of N. F. Mott}}}\ (\bibinfo  {publisher} {Springer,
  Berlin},\ \bibinfo {year} {2006})\BibitemShut {NoStop}%
\bibitem [{\citenamefont {Parisio}\ and\ \citenamefont
  {Dias}(2011)}]{ParisioDias2011}%
  \BibitemOpen
  \bibfield  {author} {\bibinfo {author} {\bibfnamefont {F.}~\bibnamefont
  {Parisio}}\ and\ \bibinfo {author} {\bibfnamefont {L.}~\bibnamefont {Dias}},\
  }\href {\doibase 10.1103/PhysRevE.84.035101} {\bibfield  {journal} {\bibinfo
  {journal} {Physical Review E}\ }\textbf {\bibinfo {volume} {84}},\ \bibinfo
  {pages} {035101} (\bibinfo {year} {2011})}\BibitemShut {NoStop}%
\bibitem [{\citenamefont {Kendal}\ and\ \citenamefont {Moran}(1963)}]{kendal}%
  \BibitemOpen
  \bibfield  {author} {\bibinfo {author} {\bibfnamefont {M.~G.}\ \bibnamefont
  {Kendal}}\ and\ \bibinfo {author} {\bibfnamefont {P.~A.~P.}\ \bibnamefont
  {Moran}},\ }\href@noop {} {\emph {\bibinfo {title} {Geometrical
  Probability}}}\ (\bibinfo  {publisher} {Charles Griffin, London},\ \bibinfo
  {year} {1963})\BibitemShut {NoStop}%
\bibitem [{\citenamefont {dos Santos}\ \emph {et~al.}(2010)\citenamefont {dos
  Santos}, \citenamefont {Barbosa}, \citenamefont {Donangelo},\ and\
  \citenamefont {Souza}}]{donangelo}%
  \BibitemOpen
  \bibfield  {author} {\bibinfo {author} {\bibfnamefont {F.~P.~M.}\
  \bibnamefont {dos Santos}}, \bibinfo {author} {\bibfnamefont {V.~C.}\
  \bibnamefont {Barbosa}}, \bibinfo {author} {\bibfnamefont {R.}~\bibnamefont
  {Donangelo}}, \ and\ \bibinfo {author} {\bibfnamefont {S.~R.}\ \bibnamefont
  {Souza}},\ }\href@noop {} {\bibfield  {journal} {\bibinfo  {journal}
  {Physical Review E}\ }\textbf {\bibinfo {volume} {81}},\ \bibinfo {pages}
  {046108} (\bibinfo {year} {2010})}\BibitemShut {NoStop}%
\bibitem [{\citenamefont {dos Santos}\ \emph {et~al.}(2011)\citenamefont {dos
  Santos}, \citenamefont {Barbosa}, \citenamefont {Donangelo},\ and\
  \citenamefont {Souza}}]{donangelo2}%
  \BibitemOpen
  \bibfield  {author} {\bibinfo {author} {\bibfnamefont {F.~P.~M.}\
  \bibnamefont {dos Santos}}, \bibinfo {author} {\bibfnamefont {V.~C.}\
  \bibnamefont {Barbosa}}, \bibinfo {author} {\bibfnamefont {R.}~\bibnamefont
  {Donangelo}}, \ and\ \bibinfo {author} {\bibfnamefont {S.~R.}\ \bibnamefont
  {Souza}},\ }\href@noop {} {\bibfield  {journal} {\bibinfo  {journal}
  {Physical Review E}\ }\textbf {\bibinfo {volume} {84}},\ \bibinfo {pages}
  {026115} (\bibinfo {year} {2011})}\BibitemShut {NoStop}%
\bibitem [{\citenamefont {Nordlund}\ \emph {et~al.}(1996)\citenamefont
  {Nordlund}, \citenamefont {Keinonen},\ and\ \citenamefont
  {Mattila}}]{nordlund}%
  \BibitemOpen
  \bibfield  {author} {\bibinfo {author} {\bibfnamefont {K.}~\bibnamefont
  {Nordlund}}, \bibinfo {author} {\bibfnamefont {J.}~\bibnamefont {Keinonen}},
  \ and\ \bibinfo {author} {\bibfnamefont {T.}~\bibnamefont {Mattila}},\
  }\href@noop {} {\bibfield  {journal} {\bibinfo  {journal} {Physical Review
  Letters}\ }\textbf {\bibinfo {volume} {77}},\ \bibinfo {pages} {699}
  (\bibinfo {year} {1996})}\BibitemShut {NoStop}%
\bibitem [{\citenamefont {Corbett}\ \emph {et~al.}(1996)\citenamefont
  {Corbett}, \citenamefont {Reid},\ and\ \citenamefont {Johnson}}]{corbett}%
  \BibitemOpen
  \bibfield  {author} {\bibinfo {author} {\bibfnamefont {G.}~\bibnamefont
  {Corbett}}, \bibinfo {author} {\bibfnamefont {S.}~\bibnamefont {Reid}}, \
  and\ \bibinfo {author} {\bibfnamefont {W.}~\bibnamefont {Johnson}},\
  }\href@noop {} {\bibfield  {journal} {\bibinfo  {journal} {International
  Journal of Impact Engineering}\ }\textbf {\bibinfo {volume} {18}},\ \bibinfo
  {pages} {141} (\bibinfo {year} {1996})}\BibitemShut {NoStop}%
\bibitem [{\citenamefont {Weaire}\ and\ \citenamefont {Rivier}(1984)}]{rivier}%
  \BibitemOpen
  \bibfield  {author} {\bibinfo {author} {\bibfnamefont {D.}~\bibnamefont
  {Weaire}}\ and\ \bibinfo {author} {\bibfnamefont {N.}~\bibnamefont
  {Rivier}},\ }\href@noop {} {\bibfield  {journal} {\bibinfo  {journal}
  {Contemporary Physics}\ }\textbf {\bibinfo {volume} {25}},\ \bibinfo {pages}
  {59} (\bibinfo {year} {1984})}\BibitemShut {NoStop}%
\bibitem [{\citenamefont {Ishii}\ and\ \citenamefont
  {Matsushita}(1992)}]{matsushita}%
  \BibitemOpen
  \bibfield  {author} {\bibinfo {author} {\bibfnamefont {T.}~\bibnamefont
  {Ishii}}\ and\ \bibinfo {author} {\bibfnamefont {M.}~\bibnamefont
  {Matsushita}},\ }\href@noop {} {\bibfield  {journal} {\bibinfo  {journal}
  {Journal of the Physical Society of Japan}\ }\textbf {\bibinfo {volume}
  {61}},\ \bibinfo {pages} {3474} (\bibinfo {year} {1992})}\BibitemShut
  {NoStop}%
\bibitem [{\citenamefont {Meibom}\ and\ \citenamefont
  {Balslev}(1996)}]{balslev}%
  \BibitemOpen
  \bibfield  {author} {\bibinfo {author} {\bibfnamefont {A.}~\bibnamefont
  {Meibom}}\ and\ \bibinfo {author} {\bibfnamefont {I.}~\bibnamefont
  {Balslev}},\ }\href@noop {} {\bibfield  {journal} {\bibinfo  {journal}
  {Physical Review Letters}\ }\textbf {\bibinfo {volume} {76}},\ \bibinfo
  {pages} {2492} (\bibinfo {year} {1996})}\BibitemShut {NoStop}%
\bibitem [{\citenamefont {Gomes}\ and\ \citenamefont
  {de~Oliveira}(2007)}]{gomes}%
  \BibitemOpen
  \bibfield  {author} {\bibinfo {author} {\bibfnamefont {M.~A.~F.}\
  \bibnamefont {Gomes}}\ and\ \bibinfo {author} {\bibfnamefont {V.~M.}\
  \bibnamefont {de~Oliveira}},\ }\href@noop {} {\bibfield  {journal} {\bibinfo
  {journal} {International Journal of Modern Physics C}\ }\textbf {\bibinfo
  {volume} {18}},\ \bibinfo {pages} {1997} (\bibinfo {year}
  {2007})}\BibitemShut {NoStop}%
\end{thebibliography}%

\end{document}